\documentclass[fleqn,twoside]{article}%
\usepackage{amsmath}
\usepackage{amsfonts}
\usepackage{amssymb}
\usepackage{graphicx}
\usepackage{cite}
\def\be{\begin{equation}}
\def\ee{\end{equation}}
\def\bi{\bibitem}
\begin{document}
\title{String cosmology in Bianchi I space-time.}
\author{A. Banerjee$^1$, Abhik Kumar Sanyal$^2$ and S. Chakraborty$^3$}
\maketitle
\noindent
\begin{center}
\noindent
$^{1,2}$ Dept of Physics, Jadavpur University, Calcutta 700 032, India.\\
$^3$ Dept of Mathematics, Jadavpur University, Calcutta 700 032, India.
\end{center}
\footnotetext{\noindent
Electronic address:\\
\noindent
$^1$ asit@juphys.ernet.in\\
$^2$ sanyal\_ ak@yahoo.com;
Present address: Dept. of Physics, Jangipur College, India - 742213.\\
$^3$ schakraborty@math.jdvu.ac.in}
\noindent
\abstract{Some cosmological solutions of massive strings are obtained in Bianchi I space-time following the techniques used by Letelier and Stachel. A class of solutions corresponds to string cosmology associated with/without a magnetic field and the other class consists of pure massive strings, obeying the Takabayashi equation of state $\rho = (1 + \omega)\lambda$.}
\maketitle
\flushbottom
\section{Introduction:}

At the early stage of tile universe a phase transition occurs as the temperature lowers below some critical temperature, and this can give rise to various topologically stable defects of which strings are of most important whose world sheets are two dimensional time-like surfaces \cite{1}. It has been noted \cite{1} that the existence of a large scale network of strings in the early universe does not contradict the present-day observations of the universe and further the vacuum strings \cite{2} can generate density fluctuations sufficient to explain the galaxy formation. \\

These strings have stress energy and they couple ,to the gravitational field so that it may be interesting to study the gravitational effects which arise from strings. \\

This has been already done by several authors, \cite{3, 4, 5}, although the general relativistic treatment of strings was pioneered by \cite{6} and \cite{7}. Letelier \cite{8} presented some cosmological solutions of massive strings in Bianchi I and Kantowski-Sachs space-time.\\

In geometrical string (massless) models, infinite number of degrees of freedom are possessed by each string for which the end points move at the speed of light. This problem is resolved by considering the realistic (massive) string model of Takabayashi \cite{9}. The energy-momentum tensor for the massive strings has been first formulated by Letelier \cite{5}, who considered the massive strings being formed by geometric strings with particles attached along its extension. Its application to general relativity first appeared in Letelier \cite{8}, although Stachel \cite{7} considered massless strings earlier. So the total energy-momentum tensor for a cloud of massive strings can be written as

\be\label{1} {T_\mu}^\nu = \rho v_\mu v^\nu - \lambda x_\mu x^\nu,\ee
where $\rho$ is the rest energy density for a cloud of strings with particles attached to them (p-strings). Thus we have

\be\label{2} \rho = \rho_p + \lambda,\ee

$\rho_p$ being the particle energy density and $\lambda$ being the string's tension density. $v^\mu$ is the four velocity for the cloud of particles and $x^\mu$ is the four vector representing the string's direction which essentially is the direction of anisotropy. Thus,

\be\label{3} v_\mu v^\mu = -1 = x_\mu x^\mu; \hspace{0.5 in}\mathrm{and}\hspace{0.5 in}v_\mu x^\mu = 0,\ee
in $(-,+,+,+)$ signature. In the present paper, we study the string cosmology in axially symmetric Bianchi-I space-time both in the presence and absence of a source-free magnetic field. The evolution of a string-dust system may be interesting in the presence of the cosmic magnetic field.\\

Melvin \cite{10} in his cosmological solution for dust and electromagnetic field argues that the presence of magnetic field is not as unrealistic as it appears to be, because for a large part of the history of evolution matter was highly ionized, and matter and field were smoothly coupled. Later during cooling as a result of expansion the ions combined to form neutral matter.\\

Since the number of unknown parameters appearing in the model exceeds the
number of field equations by one, we require one more equation to find the exact solutions. This additional equation is an assumed relation between the metric coefficients in the first case, where the string universe is associated with a magnetic field. In the second case where there is no magnetic field we have assumed an equation of state $\rho = (1 + \omega)\lambda (\omega > 0$, is a constant), which is known as Takabayashi string (or P-string) \cite{9}.\\

Since there is no direct evidence of strings in the present day universe, we are, in general, interested in constructing models of a universe that evolves purely from the era dominated by either geometric string or massive strings and ends up in a particle dominated era with or without remnants of strings.

\section{Einstein's field equations:}

We consider an axially symmetric Bianchi I metric, which is
\be\label{4} ds^2 = -dt^2 + \exp{(2\alpha)}dx^2 + \exp{(2\beta)}(dy^2+ dz^2),\ee
where,$\alpha = \alpha(t)$ and $\beta = \beta(t)$. Now, the energy-momentum tensor for the string dust with a magnetic field along the direction of the string, i.e. the $x$-direction is given by

\be\label{5} {T_{\mu}}^\nu + {E_{\mu}}^\nu = \rho v_\mu v^\nu- \lambda x_\mu x^\nu +{1\over 4\pi}\left({F_{\mu}}^\alpha{F^{\nu}}_\alpha -{1\over 4} F_{\alpha\beta}F^{\alpha\beta}{\delta_\mu}^\nu \right).\ee
In the above, ${T_{\mu}}^\nu $ is the stress-energy tensor for a string-dust system, ${E_{\mu}}^\nu $ is that for the magnetic field and $F_{\alpha\beta} $ is the electromagnetic field tensor. The other terms have already been explained in the previous section. In the co-moving co-ordinate system $v^\mu = {\delta_{0}}^\mu $ and

\be\label{6} {T_0}^0 = -\rho,~~ {T_1}^1 = -\lambda,~~ {T_2}^2 =  {T_3}^3 = 0= {T_\mu}^\nu~(\mathrm{for}~\mu \ne \nu).\ee
Further, since the magnetic field is being assumed in the $x$-direction $F_{23}$ is the only non-zero component of the electromagnetic field tensor. Maxwell equation $F_{[\mu\nu;\alpha]} = 0$ and $(F^{\mu\nu}\sqrt{-g})_{;\mu}=0$, now lead to the result (remembering that $\sqrt{-g}$ is a function of time only)

\be\label{7} F_{23} = A,\ee
$A$ being a constant quantity. So, the components of stress energy tensor for the electromagnetic field are

\be\label{8} {E_0}^0= {E_1}^1 = - {E_2}^2 = -{E_3}^3 = -{A^2\over 8\pi}\exp{(-4\beta)}.\ee
Now, choosing units such that $8\pi G = 1$, the surviving. Components of Einstein field equations

\be\label{9} {R_\mu}^\nu - {1\over 2} {\delta_\mu}^\nu R = - \big({T_\mu}^\nu + {E_\mu}^\nu\big),\ee
are

\be\label{10} 2\dot\alpha\dot\beta + \dot \beta^2 = \rho + {A^2\over 8\pi}\exp{(-4\beta)},\ee
\be\label{11} 2\ddot\beta + 3\dot\beta^2 = \lambda + {A^2\over 8\pi}\exp{(-4\beta)},\ee
\be\label{12} \ddot \alpha + \dot\alpha^2 +\ddot \beta +\dot \beta^2 + \dot \alpha\dot\beta = -{A^2\over 8\pi}\exp{(-4\beta)}.\ee
The proper volume $R^3$, expansion scalar $\theta$ and shear scalar $\sigma^2$ are respectively given by,

\be\label{13a} R^3 = \exp{(\alpha + 2\beta)},\ee
\be\label{13} \theta = {v^a}_{;a} = \dot \alpha + 2\dot\beta = 3{\dot R\over R},\ee
\be\label{14} \sigma^2 = \sigma_{\mu\nu}\sigma^{\mu\nu} = \dot\alpha^2 + 2\dot\beta^2 - {1\over 3}\theta^2,\ee
where,
\be\label{15} \sigma_{\mu\nu} = {1\over 2}\left[v_{\mu;\nu} + v_{\nu;\mu} + v_\mu v^{\alpha}v_{\nu;\alpha} + v_{\nu}v^{\alpha} v_{\mu;\alpha}\right] -{1\over 3}\theta \big(g_{\mu\nu} + v_\mu v_\nu\big),\ee
Now, one can directly obtain the Ray Chaudhuri's equation \cite{11}
from the above set of field equations \eqref{10} to \eqref{12} and using \eqref{14} and \eqref{15} as,

\be\label{16} \dot\theta = - {1\over 3}\theta^2 - 2\sigma^2 = {1\over 2}\rho_p - {A^2\over 8\pi}\exp{(-4\beta)},\ee
where,
\be\label{17} R_{\mu\nu}v^\mu v^\nu = -{\rho_p\over 2} -{A^2\over 8\pi}\exp{(-4\beta)}.\ee
Now in view of all the three (strong, weak and dominant) energy conditions \cite{12}, one finds $\rho \ge 0,~\rho_p \ge 0$, together with the fact that the sign of $\lambda$ is unrestricted, it may take values positive, negative or zero as well. This implies in view of \eqref{16} that even the existence of the strings is unable to halt the collapse. From the above energy conditions we find that $\lambda$ might even take the negative value and
therefore Einstein's equation \eqref{9} with $\lambda < 0$, is the equation for an anisotropic fluid with pressure different from zero along the direction of $x^\mu$.

\section{Exact solutions of Einstein's field equations:}

Since we have a set of three field equations \eqref{10}-\eqref{12}, with four unknown parameters, viz. $\alpha,~\beta,~\rho,~\lambda$ the reduction of one unknown enables us to obtain a set of exact solutions. This is achieved in the present case by assuming the following relation between the metric coefficients

\be\label{18} \alpha = a \beta.\ee
Where $a$ is a constant. In view of this \eqref{12}, reduces to the following form,

\be\label{18a} (a + 1) \ddot \beta + (a^2 + a + 1)\dot\beta^2 = -{A^2\over 8\pi} \exp{(-4\beta)}.\ee
For $a \ne - 1$ (note, $a = - 1$ implies $\dot\beta$ is imaginary) the above equation can be written as
an integral equation

\be\label{18b} \int d\left[\dot\beta^2 \exp{\left\{2\left({a^2 + a + 1\over a + 1}\right)\beta\right\}}\right] = -{A^2\over 4\pi(a + 1)}\int \exp{\left\{2\left({a^2 - a - 1\over a + 1}\right)\beta\right\}}d\beta + k, \ee
where $k$ is constant of integration. So one obtains

\be\label{18c} \dot\beta^2 = k \exp{\left[-2\left({a^2 + a + 1\over a + 1}\right)\beta\right]} - {A^2 \over 8\pi(a^2 -a-1)} \exp{(-4\beta)},\ee
which can again be written in an integral form as

\be\label{19}\int{e^{2\beta}d\beta\over \left[k \exp{\left\{-2\left({a^2 + a + 1\over a + 1}\right)\beta\right\}} - {A^2 \over 8\pi(a^2 -a-1)}\right]^{1\over 2}} = \pm(t-t_0),\ee
where $t_0$ is another integration constant. We shall solve the above integral for two different cases, ${(a^2 -a- 1)\over (a + 1)} = - 1~ \mathrm{and} ~2$. For other cases the solutions either will not be obtained in closed form or $`a$' will become imaginary.

\subsection{Case-1:}
In this subsection we shall deal with the case

\be\label{20} {(a^2 -a- 1)\over (a + 1)} = - 1,\ee
i.e. $a = 0$, and hence $\alpha = 0$. It gives a very special form of Bianchi-1 metric. So now, \eqref{19} can be immediately integrated to yield,

\be\label{21} \exp{(2\beta)} = k(t-t_0)^2 - {A^2\over 8\pi k}.\ee
$\rho$ and $\lambda$ can now be obtained from \eqref{10} and \eqref{11} respectively as

\be\label{22} \rho = {k\over \left[k(t-t_0)^2- {A^2\over 8\pi k}\right]},  \ee
\be\label{23} \lambda = {\left[k^2(t-t_0)^2- 3{A^2\over 8\pi}\right]\over \left[k(t-t_0)^2- {A^2\over 8\pi k}\right]^2},\ee
and hence $\rho_p$ may be expressed as.

\be\label{24} \rho_P = \rho -\lambda = {{A^2\over 4\pi}\over \left[k(t-t_0)^2- {A^2\over 8\pi k}\right]^2}.\ee
Finally, $R^3, ~\theta ~\mathrm{and}~ \sigma$ can be obtained from \eqref{13} to \eqref{16} respectively as

\be\label{25}R^3 =  k(t-t_0)^2 - {A^2\over 8\pi k},\ee

\be\label{26}\theta =  {2k(t-t_0)^2\over R^3}, \ee

\be\label{27} \sigma^2 = {1\over 6} \left[{2 k(t-t_0)\over R^3}\right]^2.\ee
From the above solutions we observe that at the initial epoch, $(t - t_0)^2 = \left({A^2\over 8\pi k^2}\right)$, the string model starts with an initial singularity $R^3 \longrightarrow 0$, while $\rho,~ \rho_p,~ \lambda,~ \theta,~ \sigma^2$ etc. diverge. This is a line singularity, since, $\exp{(2\alpha)} \longrightarrow 1$ and $\exp {(2\beta)} \longrightarrow 0$. At a later instant when $(t - t_0)^2 = {3A^2\over 8 \pi k^2}$ we have $\lambda = 0$ and $\rho = \rho_p$. So at this epoch strings vanish and we are left with a dust filled universe with a magnetic field. At this stage

\be \rho = {4\pi k^2 \over A^2};\hspace{0.3 in}R^3 = {A^2\over 4\pi k};\hspace{0.3 in}\theta = {2\over A}(6\pi)^{1\over 2};\hspace{0.3 in}\sigma^2 = {2\pi k^2\over 3A^2}.\ee
i.e. all these parameters are of finite magnitude. In this solution matter is directly related with the magnetic field as is noted in \eqref{24}. When the magnetic field is absent, the matter is also absent and the solution reduces to that of pure geometric string distribution.

\subsection{Case-2:}

In this case we shall consider,
\be\label{28} {a^2 - a -1\over a+1} =2,\hspace{0.2 in}\mathrm{i.e.}\hspace{0.2 in}a = {3\pm(21)^{1\over 2}\over 2}.\ee
For this case integral \eqref{19} can at once be evaluated to yield

\be\label{29} \exp{(2\beta)} = \left[{8\pi k\over A^2}\big\{5\pm (21)^{1\over 2}\big\} - \left({A^2 \over 2\pi}\right){(t-t_0)^2\over \big\{5\pm (21)^{1\over 2}\big\}}\right]^{1\over 2}. \ee
From solution \eqref{29} it is evident that the arbitrary constant $k$ must be positive in this case, so we replace $k$ by $m^2$ in the following. The other parameters can be obtained as before, which are:

\be\label{30} \rho = {\left[{\big\{9 + 2(21)^{1\over 2}\big\}\over \big\{5 + (21)^{1\over 2}\big\}^2}\right]\left[{A^2(t-t_0)^2\over 16\pi^2}\right] - \big\{5 + (21)^{1\over 2}\big\}m^2 \over \left[{8\pi m^2\over A^2}\big\{5\pm (21)^{1\over 2}\big\} - \left({A^2 \over 2\pi}\right){(t-t_0)^2\over \big\{5 + (21)^{1\over 2}\big\}}\right]^{2}}\ee
The positivity of $\rho$ which follows from the energy conditions demands that one has to choose the positive sign before $(21)^{1\over 2}$ in the solution \eqref{29}. With this choice the other parameters are obtained explicitly as follows,

\be\label{31}\lambda =   -{\big\{9 + (21)^{1\over 2}\big\}m^2-\left[{\big\{4 + (21)^{1\over 2}\big\}\over \big\{5 + (21)^{1\over 2}\big\}^2}\right]\left[{A^2(t-t_0)^2\over 16\pi^2}\right]\over \left[{8\pi m^2\over A^2}\big\{5\pm (21)^{1\over 2}\big\} - \left({A^2 \over 2\pi}\right){(t-t_0)^2\over \big\{5  + (21)^{1\over 2}\big\}}\right]^2}\ee

\be\label{32}\rho_p = {4m^2 + {A^4(t-t_0)^2\over 16\pi^2\big\{5  + (21)^{1\over 2}\big\}}\over \left[{8\pi m^2\over A^2}\big\{5\pm (21)^{1\over 2}\big\} - \left({A^2 \over 2\pi}\right){(t-t_0)^2\over \big\{5 + (21)^{1\over 2}\big\}}\right]^2}\ee

\be\label{33}R^3 = \left[{8\pi m^2\over A^2}\big\{5\pm (21)^{1\over 2}\big\} - \left({A^2 \over 2\pi}\right){(t-t_0)^2\over \big\{5 + (21)^{1\over 2}\big\}}\right]\times {\big\{7 + (21)^{1\over 2}\big\}\over 8} \ee

\be\label{34} \theta = \left[{\big\{7 + (21)^{1\over 2}\big\}\over \big\{5 + (21)^{1\over 2}\big\}}\right]\left[{A^2\over 8\pi}\right]\left[{t-t_0\over {8\pi m^2\over A^2}\big\{5\pm (21)^{1\over 2}\big\} - \left({A^2 \over 2\pi}\right){(t-t_0)^2\over \big\{5 + (21)^{1\over 2}\big\}}}\right]\ee

\be\label{35}\sigma^2 = \left[{\big\{7 + (21)^{1\over 2}\big\}\over \big\{5 + (21)^{1\over 2}\big\}^2}\right]\left[{A^2\over 48\pi^2}\right]\left[{(t-t_0)\over {8\pi m^2\over A^2}\big\{5\pm (21)^{1\over 2}\big\} - \left({A^2 \over 2\pi}\right){(t-t_0)^2\over \big\{5 + (21)^{1\over 2}\big\}}}\right]\ee
So, we get the complete set of solutions. Now, it is evident from the solution \eqref{29} that

\be U^2 < {16\pi^2 m^2 \over A^4}\big\{5  + (21)^{1\over 2}\big\}^2,\ee
where $U = t_0 - t$, so when $t < t_0,$ we have $U > 0$, and $\theta > 0$, i.e. the expanding model. Towards the past $U^2$ increases and at an epoch $t = - t_s$, we find

\be U_s^2 = (t_s+t_0)^2 = {16\pi^2 \over A^4}\big\{5  + (21)^{1\over 2}\big\}^2,\ee
which shows the existence of a singularity in the finite past. At $t$ increases, $U^2$ decreases and we have the regularity condition satisfied when $t$ approaches $t_0$ that is $U^2 \rightarrow 0$. This gives the upper limit to the expansion, as is evident from \eqref{34}. At a subsequent
stage the motion reverses from a maximum volume and $t > t_0$ so that the system contracts $(\theta < 0)$. As $t$ increases, evidently $U^2$ increases and approaches the singularity when $t = t_s + 2t_0$.\\

It should be clearly mentioned in this context that, the field equations without any contribution from the magnetic field can however, be integrated to yield solutions, which are not special cases of those given above.\\

In accordance with the assumption \eqref{18}, the field equation \eqref{12} would reduce to the following form

\be\label{36} \ddot \beta + K\dot\beta^2 = 0, \hspace{0.1 in}\mathrm{where},\hspace{0.1in} K = {a^2 + a + 1\over a+1},\ee
which admits the following general solution:
\be\label{37} \exp{(2\beta) = \big[C (t-t_0)\big]^{2\over K}},\ee
where $C$ is a constant. Thus,
\be\label{38} \exp{(2\alpha) = \big[C (t-t_0)\big]^{2a\over K}}.\ee
Hence, all other physical parameters can be obtained in a straightforward manner as,
\be\label{39} R^3 = \big[C (t-t_0)\big]^{2 + a\over K},\ee
\be\label{40} \rho = {2a + 1\over (t-t_0)^2}\ee
\be\label{41} \lambda = {1\over (t-t_0)^2}\ee
\be\label{42} \rho_p = {2a \over (t-t_0)^2}\ee
\be\label{43} \theta = {a + 2\over (t-t_0)}\ee
\be\label{44} \sigma^2 = {2\over 3}\left[{a - 1\over t-t_0}\right]^2\ee
From the above solutions, we note that at the initial epoch, $t = t_0$, $R^3 \rightarrow 0$, while
$\rho,~ \lambda,~ \rho_p,~ \theta,~ \sigma^2$ etc diverge, and both $\exp{(2\alpha)} \rightarrow 0$, $\exp{(2\beta)}\rightarrow 0$, exhibiting the point singularity. This is the starting point of the string model. Again at a later stage, when $t \rightarrow \infty$, $R^3$ also tends to infinity but all other physical parameters become insignificant. It is interesting to
note that for a pure geometric string $(\rho_p = 0)$ model, we have to take the value of $a = 0$. For this case the universe starts with strings and ends up at a stage when the massive strings themselves disappear without any remnant.

\section{P-string solutions in the absence of magnetic field:}

In this section we consider a string-dust system in the absence of the magnetic field, in which case the field equations take the following form,

\be\label{45} 2\dot\alpha \dot \beta + \dot \beta^2 = \rho,\ee
\be\label{46} 2\ddot\beta + 3 \dot \beta^2 = \lambda,\ee
\be\label{47} \ddot\alpha + \ddot\beta + \dot\alpha^2 + \dot\beta^2 + \dot\alpha\dot\beta = 0.\ee
Here again, since we have to deal with four unknown parameters being involved in a set of three equations, so in order to obtain the exact solutions of the above set of field equations, we require one more equation. Let that be an equation of state for Takabayashi string (i.e. P-string), which is

\be\label{48} \rho = (1 + \omega)\lambda,\ee
where $\omega > 0$ is a constant. This is analogous to choosing an equation of state between the matter density and the string's tension density in the form $\rho_p = \omega \cdot\lambda$ This equation of state for Takabayashi string is interesting for the two limiting cases viz. for infinitesimally small value of $\omega$ on the one hand and for infinitely large value of $\omega$ on the other. When $\omega$ is very small only geometric strings appear and for very large $\omega$, particles dominate over strings.\\

In view of \eqref{48}, one can reduce the set of equations \eqref{45} to \eqref{47} into a single nonlinear differential equation in $\beta$, as

\be\label{49} {\ddot\beta\over \beta} + \omega{\dot\beta^2\over \beta^2} + \left({6\omega^2 + 15 \omega + 10\over 2(\omega + 1)}\right)\ddot\beta + {3\over 4} \left({3\omega^2 + 6 \omega + 4\over \omega + 1}\right)\dot\beta^2 = 0.\ee
This equation can be solved with a special technique of substitutions which is as follows. We first replace $\dot\beta^2$ by $y$ and then substitute $z$ for $y_\beta^2 . y^\omega$ and so the above equation \eqref{49} becomes

\be\label{50} y^{-{\omega\over 2}}.z_y + k\sqrt z + l\cdot y^{\big({\omega\over 2} + 1\big)}= 0,\ee
where the constants $k$ and $l$ are:

\be\label{51} k = {6\omega^2 + 15 \omega + 10\over (\omega + 1)},\hspace{0.3 in} \mathrm{and}\hspace{0.3 in} l = 3\left[{3\omega^2 + 6 \omega + 4\over \omega + 1}\right].\ee
Now, again replacing $y^{({\omega\over 2}+1)}$ by $U$ and $z$ by $V^2$ and considering $V = AU$, $A$ being
a constant, \eqref{50} takes the following form:

\be (\omega + 2) A + k + {l\over A} = 0,\ee
which immediately yields the value of the constant $A$, which is

\be\label{52} A = -3, \hspace{0.3 in}\mathrm{or},\hspace{0.3 in} A = - \left[{3\omega^2 + 6 \omega + 4\over (\omega + 1)(\omega+2)}\right],\ee
$k$ and $l$ being substituted from \eqref{51}. So now one obtains

\be A = {V\over U} = - {\sqrt z\over y^{({\omega\over 2}+ 1)}} = {Y_\beta\over Y},\ee
and hence,

\be \ln{Y}  = A \cdot \beta + m,\ee
where $m$ is a constant of integration. Now replacing $Y$ by $\dot\beta^2$ the above equation can be solved to obtain

\be\label{53} \exp{(2\beta)} = \exp{\left(-{2m\over A}\right)}\left[-{2\over A(t-t_0)}\right]^{4\over A}.\ee
In view of \eqref{53} one can easily find out the other parameters from \eqref{45} to \eqref{47} as

\be\label{54} \exp{2\alpha} = n(t-t_0)^{-2\left[{A(1+\omega)+3\omega+2\over A}\right]},\ee
\be\label{55} \rho = {4\over A^2}(\omega+1)(A+3)(t-t_0)^{-2},\ee
\be\label{56}\lambda = {4\over A^2}(A+3)(t-t_0)^{-2}\ee
\be\label{57}\rho_p = {4\over A^2}\omega(A+3)(t-t_0)^{-2}\ee
where $n$ is yet another constant of integration. Finally, $R^3,~ \theta~ \mathrm{and}~ \sigma^2$ can be obtained from \eqref{13} to \eqref{15} as

\be\label{58} R^3 = \sqrt{n} \left(-{2\over A}\right)^{4\over A} \exp{\left(-{2m\over A}\right)} (t-t_0)^{\left[-{A(1+\omega)+3(2+\omega)\over A}\right]}, \ee
\be\label{59}\theta = -\left[{A(1+\omega)+3(2+\omega)\over A}\right](t-t_0)^{-1},\ee
\be\label{60} \sigma^2 = {2\over 3}\left[{A(1+\omega)+3\omega\over A}\right]^2(t-t_0)^{-2}.\ee
Now for $A = - 3$, $\rho,~, \lambda,~ \rho_p$ all become zero and we are left with,

\be\begin{split}\label{61}& \exp{(2\beta)} = \exp{\left(2m\over 3\right)}\left({2\over 3(t-t_0)}\right)^{-{4\over 3}}= m_1(t-t_0)^{4\over 3},\\&
\exp{(2\alpha)} = n (t-t_0)^{-{2\over 3}},\\&
R^3 = \sqrt{n} {\left(2\over 3\right)}^{-{4\over 3}}\exp{\left(2m\over 3\right)}(t-t_0) = l_1(t-t_0),\\&
\theta = (t-t_0)^{-1},\\&
\sigma^2 = {2\over 3}(t-t_0),\end{split}\ee
where, $m_l$, $l_1$ are constants depending on $m$ and $n$. Now absorbing the constants $m_l$ and $n$ into the metric coefficients by suitable choice of spatial co-ordinates and defining a new time co-ordinate $t =(t - t_0)$, it is possible to write the metric \eqref{21} in the following form,

\be ds^2 = -d\bar t^2 + \bar t^{2p_1}dx^2 + \bar t^{2p_2}dy^2+ \bar t^{2p_3}dz^2, \ee
where,

\be p_1 = -{1\over 3},\hspace{0.2 in}p_2 = p_3 = {2\over 3}.\ee
From the above it is evident that

\be p_l + p_2 +p_3 = 1 = p_1^2 +p_2^2 +p_3^2.\ee
So this metric is the well-known Kasner metric for a flat, homogeneous anisotropic empty space. It may be noted that Letelier \cite{8} could not find any solution, although he concluded that the anisotropy of the universe decreases as the universe expands. Again for

\be A = -\left[{3\omega^2+ 6\omega + 4\over (\omega+1)(\omega+2)}\right] = -\mu~(\mathrm{say}),\ee
the solutions take the following forms:

\be\begin{split}\label{62} & \exp{(2\beta)} = m_2 (t-t_0)^{4\over \mu},\\&
\exp{(2\alpha)} = n(t-t_0)^{\left[4n\omega\over \mu(\omega+2)\right]},\\&
\rho = {4(3\omega+2)(\omega+1)^2(\omega+2)\over (3\omega^2 + 6\omega + 4)^2}(t-t_0)^{-2},\\&
\lambda = {4(3\omega+2)\over \mu(3\omega^2 + 6\omega + 4)^2}(t-t_0)^{-2},\\&
\rho_p = {4\omega(3\omega+2)\over \mu(3\omega^2 + 6\omega + 4)}(t-t_0)^{-2},\\&
R^3 = l_2 (t-t_0)^{\left[3\omega + 4\over \mu(\omega+2)\right]},\\&
\theta = {3\omega + 4\over \mu(\omega + 2)}(t-t_0)^{-1},\\&
\sigma^2 = \left[8\over 3\mu^2(\omega+2)^2\right](t-t_0)^{2},
\end{split}\ee
where $m_2$ and $l_2$ are constants depending on $m$, $n$ and $w$.
So \eqref{62} is the set of solutions obtained for Takabayashi string. Here we observe that at the initial epoch when $t \rightarrow t_0$, $\exp{(2\alpha)},~ \exp{(2\beta)}$ and hence $R^3$ all tend to zero giving rise to a point singularity, and all other parameters viz., $\rho,~\rho_p,~\lambda,~\theta,~\sigma^2$, etc. diverge. As $t \rightarrow \infty$, $\exp{(2\alpha)}\rightarrow \infty,~ \exp{(2\beta)}\rightarrow \infty$ and hence $R^3\rightarrow \infty$, while all other parameters become insignificant.

\section{Conclusion:}
The present work extends the work of Letelier \cite{8} given earlier. In addition, we have considered a source-free magnetic field, the reason for which has been discussed in the introduction. We have obtained three sets of solutions with a special choice of metric coefficients, viz., $\alpha = a\beta$. For a particular value of $a$ we observe that our model starts from a string dominated era, but at a later instant strings vanish and the universe becomes particle dominated. Here, the gravitational field is coupled to the magnetic field such that in the absence of the magnetic field the system reduces to pure geometric strings. For a different value of $a$ we find that there is an upper limit for the expansion of the string universe and one cannot obtain a physical model in the absence of the magnetic field. We have also presented solutions for vanishing magnetic field and it is observed that the string distribution cannot go over to a particle dominate era at any stage of the evolution, unlike the result obtained by Letelier \cite{8}. At the end an explicit solution for Takabayashi strings \cite{9} has been presented
which represents a system of massive strings.

\end{document}